\documentclass[a4paper,reqno,12pt]{amsart}
\usepackage{amssymb,euscript,bbold}
\usepackage{array}
\renewcommand{\d}{\partial}

\renewcommand{\gg}{\mathfrak g}

\newcommand{\D}{\mathsf D}
\newcommand{\sT}{\mathsf T}

\newcommand{\1}{\mathbb 1}

\newcommand{\Z}{\mathbb Z}

\newcommand{\AdS}{\text{AdS}}

\newcommand{\SU}{\mathrm{SU}}

\newcommand{\eF}{\EuScript{F}}

\DeclareMathOperator{\Ad}{Ad}
\DeclareMathOperator{\Tr}{Tr}
\begin{document}

\title[]{D-branes in group manifolds} 
\author[]{Sonia Stanciu}
\address[]{\begin{flushright}Theoretical Physics Group\\
Blackett Laboratory\\
Imperial College\\
London SW7 2BZ, UK\\
\end{flushright}}
\email{s.stanciu@ic.ac.uk}
\date{\today}
\thanks{Imperial/TP/98-99/59}

\begin{abstract}
In this paper we re-examine the geometric interpretation of gluing
conditions in WZW models and the possible D-brane configurations that
they give rise to.  We show how the boundary conditions are encoded in
the gluing conditions imposed on the chiral currents.  We analyse two
special classes of gluing conditions: the first, which preserves the
affine symmetry of the bulk theory, describes D-branes whose
worldvolumes are given by `twisted' conjugacy classes; the second
class describes configurations which include subgroups and cosets.
\end{abstract}
\maketitle

\tableofcontents

\section{Introduction}

D-branes in group manifolds have received recently a great deal of
interest, as they provide an ideal laboratory for the study of the
intricacies related to the description of D-branes in general curved
backgrounds.  There is already an extensive amount of literature
devoted to the analysis of D-branes in exact string backgrounds (see,
for instance, \cite{OOY,BBMOOY,KO,SDKS,RS,FuSch,STDNW,AS}), where by
using the methods of CFT one can obtain a microscopic description of
various D-brane configurations in terms of conformally invariant
boundary states and/or boundary conditions.  However the geometric
interpretation of these configurations proves to be a rather difficult
problem, a fact reflected, in particular, in a number of partially
contradicting results which appeared in the past few years
(e.g. \cite{KO,STDNW,AS,SAdS3}).  This has generated a certain
controversy in the literature surrounding the very definition and
meaning of the Neumann and Dirichlet boundary conditions in curved
backgrounds, particularly in the case of a group manifold.

The aim of these notes is twofold.  On the one hand, we describe an
improved approach to the geometric interpretation of D-branes
configurations in group manifolds obtained via the boundary state
approach; in this sense, this work is a direct continuation of
\cite{SDKS,STDNW,SAdS3}.  Some of the results which we derive here
were already used in \cite{SAdS3} to determine the possible D-branes
in the type IIB string background on $\AdS_3\times S^3\times T^4$ with
an NS B field.  Here, however, we describe this method in detail and
in as much generality as possible at the moment.  On the other hand,
in the process of doing this, we hope to clarify some discrepancies
that can be found in the literature.  In order to do this we find it
useful to start by reviewing, in Section 2, the well-understood case
of the free bosonic string and then work our way up to the more subtle
case of group manifolds.  Also, in order to be able to directly
compare different results we will work, when necessary, both in the
closed and in the open string pictures.

The central question in the geometric interpretation of a D-brane
configuration obtained via the boundary state approach is finding the
correct generalisation of the Neumann and Dirichlet boundary
conditions of the free case.  As is well-known, boundary conditions in
a free bosonic theory are of the form
\begin{equation*}
\d X^{\mu}(z) = {R^{\mu}}_{\nu} \bar\d X^{\nu}(\bar z)~.
\end{equation*}
where the information about the Neumann and Dirichlet directions is
encoded in the eigenvalues and eigenvectors of the constant matrix
$R$.  On the other hand, at the quantum level, the above conditions
are to be understood as gluing conditions imposed on the chiral
currents $\d X(z)$ and $\bar\d X(\bar z)$, which are the free fields
in terms of which the corresponding conformal field theory is defined.
These conditions thus play a crucial role in controlling the conformal
invariance of the resulting D-brane configuration.

This type of conditions can further be applied to the study of
boundary states in flat backgrounds, or in Calabi-Yau manifolds
\cite{OOY,BBMOOY}.  In this latter case it has been shown that the
geometry of the corresponding D-branes, in the large volume limit, is
encoded in the matrix of boundary conditions.

If we want to study D-branes in curved backgrounds we are faced with
two possible avenues (assuming that we are dealing with backgrounds
for which we know both the CFT and the corresponding sigma model).  We
can try to impose gluing conditions on the chiral fields in terms of
which the CFT is defined (and then interpret them geometrically) or,
alternatively, we can derive the boundary conditions from the sigma
model action (and then impose the requirement of conformal
invariance).  However, since the basic requirement of the boundary
state approach is the conformal invariance\footnote{This is not the
only condition one must impose in order to get obtain a physically
acceptable D-brane spectrum, as it was pointed out recently in
\cite{HKMS}.  Here, however, we will see that at least in the case of
group manifolds, already the conditions imposed independent of a
detailed knowledge of the spectrum of the quantum theory restrict
significantly the possible D-brane configurations (that is, their
geometry).} of the resulting D-brane configuration, it appears natural
to generalise the above gluing conditions in such a way that
conformal invariance remains under control.

In the particular case of strings moving on a group manifold, which
are described by the WZW model, the fields in terms of which the
conformal field theory is realised are the affine currents.  Therefore
the natural nonabelian generalisation of the gluing conditions of the
free bosonic string takes the form
\begin{equation*}
J_a (z) = {R^b}_a \bar J_b (\bar z)~.
\end{equation*}
This approach, which is described in Section 3, does however create a
problem when it comes to the geometric interpretation of the solutions
to these gluing conditions.  The reason is the following.  In the flat
space case the gluing conditions on the free fields $\d X^{\mu}$ and
$\bar\d X^{\mu}$ are, at the same time, boundary conditions taking
values in the tangent space of the target manifold at a given but
otherwise arbitrary point.  By contrast, the nonabelian gluing
conditions on the chiral currents of the WZW theory are \emph{not},
strictly speaking, boundary conditions; rather, since the currents
themselves are Lie algebra-valued objects, they take values in the
tangent space to the group manifold \emph{at the identity}.  It
therefore follows that in order to interpret geometrically the
algebraic gluing conditions imposed on the affine currents we must
first of all `translate' them into boundary conditions.  In Section 4
we show how this is achieved.  It turns out that, unlike the flat
space case, the boundary conditions in a group manifold are point
dependent, and this fact is essentially due to the nonabelian nature
of the target manifold.

The simplest and, in some sense, the most natural class of gluing
conditions is the one characterised by the fact that it preserves the
infinite-dimensional symmetry of the current algebra.  In Section 5 we 
analyse in detail this type of gluing conditions, which we will refer
to as type-D.  There we show that type-D gluing conditions defined by
an inner automorphism describe odd-dimensional D-branes whose
worldvolumes are conjugacy classes translated by a certain group
element.  In the most general case, the D-branes worldvolumes describe 
a generalised version of conjugacy classes which we briefly discuss.

In Section 6 we analyse a different yet closely related type of gluing
conditions, which we call type-N.  They do not preserve the current
algebra of the bulk theory and, generically, they give rise to more
complicated D-brane configurations for which we lack, at the moment, a
full picture.  However we are able to discuss in detail a particular
example, for which we identify solutions including subgroups, cosets
and open submanifolds of the same dimension as the target manifold.

Finally, in the Appendix we collect a few useful facts about (twisted)
conjugacy classes.

\section*{Note added in proof}
While this paper was in its final stages a paper \cite{FFFS} appeared
where the same problem of D-branes in WZW models is analysed.  While
there is a certain overlap between some of the results, the two papers
use significantly different techniques, and the exact relation between
them still remains to be clarified.

\section{Boundary conditions and boundary states in flat space} 

The aim of this section is to review a few well-known facts about
D-branes in a flat string background, in order to set the stage for
the discussion of the group manifold case later on.

D-branes can be studied in a variety of ways: by using the techniques
of perturbative string theory, they can be described either in terms
of boundary conditions of open strings or as boundary states in the
closed string sector.  Boundary conditions were initially derived from
the vanishing of the boundary term in the action of the free string
(or, more generally, in the case of a flat background).  A Neumann
boundary condition imposed in a certain direction ensures that there
is no flow of momentum through the boundary of the string worldsheet,
that is
\begin{equation*}
\left. \d_n X^{||} \right|_{\d\Sigma} = 0~.
\end{equation*}
Here, the subscript $n$ indicates the direction normal to the boundary
of the worldsheet $\d\Sigma$, whereas the superscript $||$ denotes the
direction parallel to the D-brane worldvolume.  By contrast, a
Dirichlet boundary condition in a given direction constrains the
boundary of the string worldsheet to lie within the hyperplane normal
to that direction.  This implies the following condition
\begin{equation*}
\left. \d_t X^{\perp} \right|_{\d\Sigma} = 0~,
\end{equation*}
where the subscript $t$ refers to the direction tangent to the
boundary of the string worldsheet and $\perp$ denotes the component of 
the background field normal to the worldvolume of the D-brane.
 
For what follows it will prove useful to write down the Neumann and
Dirichlet conditions explicitly, both in the closed and in the open
string pictures.  For this we need the mode expansion of the free
bosonic string
\begin{equation*}
X^{\mu}(\sigma,\tau) = X^{\mu}_L(\tau+\sigma) +
                       X^{\mu}_R(\tau-\sigma)~, 
\end{equation*}
where
\begin{equation*}
\d_+ X^{\mu}_L = \sum_{n\in\Z}\alpha^{\mu}_n~
                 e^{-in(\tau+\sigma)}~,\qquad 
\d_- X^{\mu}_R = \sum_{n\in\Z} \tilde\alpha^{\mu}_n~
                 e^{-in(\tau-\sigma)}~, 
\end{equation*}
with the standard notation $\d_{\pm} = \d_{\tau}\pm\d_{\sigma}$.  The
canonical commutation relations for the modes read
\begin{align}\label{eq:oscalg}
[\alpha^{\mu}_n,\alpha^{\nu}_m] &=
n~\delta_{m+n,0}~\eta^{\mu\nu}~,\nonumber\\
[\tilde\alpha^{\mu}_n,\tilde\alpha^{\nu}_m] &= n~
                                     \delta_{m+n,0}~\eta^{\mu\nu}~,\\ 
[\alpha^{\mu}_n,\tilde\alpha^{\nu}_m] &= 0~.\nonumber
\end{align}

Let us now consider the open and the closed string pictures
separately.

\subsection*{Closed string}

The boundary of the worldsheet in this case is taken to be at $\tau =
0$.  A Neumann boundary condition in a given direction say, $X^{\mu}$,
is therefore given by the condition
\begin{equation*}
\left. \d_{\tau} X^{\mu}\right|_{\tau=0} = 0~.
\end{equation*}
By contrast, a Dirichlet boundary condition in the same direction
reads 
\begin{equation*}
\left. \d_{\sigma} X^{\mu}\right|_{\tau=0} = 0~.
\end{equation*}
One can easily work out these conditions in terms of the modes and
obtain
\begin{align*}
\alpha^{\mu}_n + \tilde\alpha^{\mu}_{-n} &= 0~,\tag{N}\\
\alpha^{\mu}_n - \tilde\alpha^{\mu}_{-n} &= 0~.\tag{D}
\end{align*}

Notice that the above conditions on the modes are not to be taken as
operatorial relations on the quantum fields.  In fact, one can
easily check that they do \emph{not} yield automorphisms of the
oscillator algebra \eqref{eq:oscalg}.  Rather, they are only satisfied
on certain \emph{boundary states} $|B\rangle$, which are characterised
by the very property that they satisfy a certain boundary condition,
that is
\begin{align*}
(\alpha^{\mu}_n + \tilde\alpha^{\mu}_{-n})|B\rangle &= 0~,\tag{N}\\
(\alpha^{\mu}_n - \tilde\alpha^{\mu}_{-n})|B\rangle &= 0~,\tag{D}
\end{align*}
in a given direction $X^{\mu}$.  

A boundary state $|B\rangle$ is specified by a complete set of
boundary conditions, for all the directions of the target manifold.  A
convenient way of encoding this information is to introduce a set of
linear operators $\{\phi^{\mu}_n\}$ which annihilate $|B\rangle$; they
are given by $\phi^{\mu}_n = \alpha^{\mu}_n \pm
\tilde\alpha^{\mu}_{-n}$, depending on whether we have a Neumann or a
Dirichlet boundary condition in the direction $X^{\mu}$.
Consistency then requires that a boundary state which is annihilated
by two operators, say $\phi^{\mu}_n$ and $\phi^{\nu}_m$, must also be
annihilated by their commutator, that is
\begin{equation*}
[\phi^{\mu}_n,~\phi^{\nu}_m]|B\rangle = 0~.
\end{equation*}
One can easily check that this is always satisfied, in other words, we 
do not have any restriction on the choice of Neumann and Dirichlet
conditions that we impose.

Since in practice we usually work in light-cone coordinates on the
cylinder or on the complex plane, it is useful to write down the
Neumann and Dirichlet boundary conditions in all these coordinates.
This is a straightforward exercise, and the results are summarised in
the left column of Table~\ref{tab:freeND}.  Notice in particular that
if we pass from the cylinder worldsheet to the annulus in the
complex plane, by defining $z = e^{i(\tau+\sigma)}$ and $\bar z =
e^{i(\tau-\sigma)}$, then the boundary of the worldsheet is at $z\bar
z =1$. 

The most important requirement that a boundary state must satisfy is
conformal invariance.  In the case of a bosonic theory this means that
the holomorphic and antiholomorphic sectors satisfy the following
condition at the boundary
\begin{equation*}\label{eq:ci}
\left(\sT (z) - \bar\sT (\bar z)\right)|B\rangle = 0~,
\end{equation*}
where, in this case, $\sT = \d X\cdot\d X$.  It is easy to see
that any boundary state characterised by an arbitrary combination of
Neumann and Dirichlet conditions does satisfy the above condition and
therefore preserves conformal invariance.

This set-up can be slightly generalised by introducing a matrix $R$
defining a boundary state of the form 
\begin{equation}\label{eq:cBalpha}
(\alpha^{\mu}_n + {R^{\mu}}_{\nu}\tilde\alpha^{\nu}_{-n})|B\rangle = 0~.
\end{equation}
In this case the linear operators which annihilate $|B\rangle$ are
given by $\phi^{\mu}_n = \alpha^{\mu}_n +
{R^{\mu}}_{\nu}\tilde\alpha^{\nu}_{-n}$, and the corresponding
consistency conditions imply that the matrix $R$ must preserve the
flat metric of the target space
\begin{equation*}
R^T\eta R = \eta~.
\end{equation*}

\begin{table}[h!]
\renewcommand{\arraystretch}{2.0}
\begin{tabular}{|>{$}c<{$}|>{$}c<{$}|}
\hline
\text{CLOSED STRING} & \text{OPEN STRING}\\ 
\hline\hline
\left. \d_{\tau} X^{\mu}\right|_{\tau=0} = 0 & 
       \left. \d_{\sigma} X^{\mu}\right|_{\sigma=0} = 0 \\
\left. \d_{\sigma} X^{\mu}\right|_{\tau=0} = 0 & 
       \left. \d_{\tau} X^{\mu}\right|_{\sigma=0} = 0 \\
\hline
(\alpha^{\mu}_n \pm \tilde\alpha^{\mu}_{-n})|B\rangle = 0 & 
\alpha^{\mu}_n \mp \tilde\alpha^{\mu}_{n} = 0 \\
\hline
(\d_+ X^{\mu} \pm \d_- X^{\mu})|B\rangle = 0 & 
\left.(\d_+ X^{\mu} \mp \d_- X^{\mu})\right|_{\sigma=0} = 0 \\
\hline
(z^2 \d X^{\mu} \pm \bar\d X^{\mu})|B\rangle = 0 & 
\left.(\d X^{\mu} \mp \bar\d X^{\mu})\right|_{z=\bar z} = 0\\
\hline
(\d X^{\mu}dz \mp \bar\d X^{\mu}d\bar z)|B\rangle = 0 & 
\left.(\d X^{\mu}dz \mp \bar\d X^{\mu}d\bar z)\right|_{z=\bar z} = 0\\
\hline
\end{tabular}
\vspace{8pt}
\caption{The Neumann and Dirichlet boundary conditions for free
  fields.(In each row, the first relation is a Neumann condition,
  whereas the second one is a Dirichlet condition.)\label{tab:freeND}}  
\end{table}
 
\subsection*{Open string}

The boundary of the string worldsheet it is now at $\sigma = 0$.  The
Neumann boundary condition in the direction $X^{\mu}$ is therefore
given now by
\begin{equation*}
\left. \d_{\sigma} X^{\mu}\right|_{\sigma=0} = 0~,
\end{equation*}
whereas the Dirichlet boundary condition reads
\begin{equation*}
\left. \d_{\tau} X^{\mu}\right|_{\sigma=0} = 0~.
\end{equation*}
Once again, we can write these conditions in terms of the modes, thus
obtaining
\begin{align*}
\alpha^{\mu}_n - \tilde\alpha^{\mu}_{n} &= 0~,\tag{N}\\
\alpha^{\mu}_n + \tilde\alpha^{\mu}_{n} &= 0~.\tag{D}
\end{align*}
However in this case, by contrast to the closed string picture,
both the Neumann and Dirichlet boundary conditions preserve the
algebra of the modes.

For the sake of completeness, one can write the Neumann and Dirichlet
boundary conditions in light-cone coordinates or in the complex plane,
where the boundary of the worldsheet is this time at $z=\bar z$.  The
results can be found in the right column of Table~\ref{tab:freeND}.

A short glance at the table reveals the basic rule of thumb that any
particular boundary condition in the closed string picture differs by
a relative minus sign from the corresponding open string condition.
Since this may prove rather confusing in practice, it is maybe worth
keeping in mind that the Neumann and Dirichlet boundary conditions,
written in terms of the invariant geometric objects, $\d X dz$ and
$\bar\d X d\bar z$, have the same form in the closed and open string
cases.

Also in this case we can define generalised boundary conditions for
the open string by introducing a matrix $R$ satisfying
\begin{equation}\label{eq:flatspaceBC}
\left.(\d X^{\mu} - {R^{\mu}}_{\nu}\bar\d X^{\nu})\right|_{z=\bar z} = 
                                                                  0~. 
\end{equation}
The requirement of conformal invariance implies that the matrix $R$
must preserve the metric $\eta$ of the target manifold.  Moreover, we
have the option of imposing that the free field algebra be preserved;
however in this case we do not obtain any additional condition on $R$.

\section{Gluing conditions and boundary states in group manifolds} 

Strings moving on a group manifold are described using the WZW model
\cite{GeW,KZ}, which is a solvable theory, much in the same way as a
free theory is.  The data necessary to describe such an exact string
background is a Lie group $\mathbf{G}$ together with an bi-invariant
metric, in terms of which the corresponding WZW action is given by
\begin{equation*}
I[g] = \int_{\Sigma} \langle g^{-1}\d g, g^{-1}\bar\d g\rangle +
       {\textstyle \frac{1}{6} }\int_B \langle g^{-1}dg,
       [g^{-1}dg,g^{-1}dg]\rangle~,
\end{equation*}
where the field $g$ is a map from a closed orientable Riemann surface
$\Sigma$ to the group $\mathbf{G}$.  We denote by $\gg$ the
corresponding Lie algebra, and we choose for it a basis of generators
$\{T_a\}$.  The invariant metric on $\gg$ has components $G_{ab}\equiv
\langle T_a,T_b\rangle$.  The exact conformal invariance of this model
is based, as is well known, on its infinite-dimensional symmetry group
$\mathbf{G}(z)\times\mathbf{G}(\bar z)$ characterised by the conserved
currents 
\begin{equation}\label{eq:cc}
J(z)=-\d g g^{-1}~,\qquad\qquad \bar J(\bar z)=g^{-1}\bar\d g~,
\end{equation}
which are the natural dynamical variable in this case.  If we
introduce the mode expansions for the two currents
\begin{equation*}
J_a(z) = \sum_{n\in\Z}~J_{an}~z^{-n-1}~,\qquad\qquad
\bar J_a(\bar z) = \sum_{n\in\Z}~\bar J_{an}~\bar z^{-n-1}~,
\end{equation*}
we can write down the affine Lie algebra that they satisfy
\begin{align}\label{eq:affalg}
[J_{an},J_{bm}] &= {f_{ab}}^c J_{c~n+m} +
                    n~\delta_{n+m,0}~G_{ab}~,\nonumber\\ 
[\bar J_{an},\bar J_{bm}] &= {f_{ab}}^c \bar J_{c~n+m} +
                             n~\delta_{n+m,0}G_{ab}~,\\ 
[J_{an},\bar J_{bm}] &= 0~.\nonumber
\end{align}
It is important to stress that the relative sign in the definition
\eqref{eq:cc} of the two chiral currents is \emph{not} optional: this
definition ensures that the two currents satisfy the same chiral
algebra.

The corresponding CFT is then described by the energy-momentum tensor 
\begin{equation}\label{eq:GT}
\sT(z) = \Omega^{ab}(J_a J_b)(z)~, 
\end{equation}
where $\Omega^{ab}$ are components of the inverse of the following
invariant metric with the components given by $\Omega=2G + \kappa$,
where $\kappa$ is the Killing form of $\gg$ \cite{Msug,FSSug,FS3}.
The central charge of this CFT is given by $c = \dim\gg -
\Omega^{ab}\kappa_{ab}$.  In the particular case of a simple Lie group
(algebra), the metric $\Omega$ is a multiple of the Killing form, say
$\Omega = \mu \kappa$.  The usual formulas are then recovered by
taking $\mu = (x + 2g^*)/2g^*$, with $g^*$ the dual Coxeter number and
$x$ the level.

So far most attempts to the study of D-branes in group manifolds and
coset spaces have tried to exploit the similarities of the WZW model
with a free theory (the solvability, the infinite-dimensional
symmetry, the exact conformal invariance), in order to construct some
kind of nonabelian generalisation of the boundary conditions and
boundary states in flat space.  In other words, one treats the group
manifold case as a generalisation of the flat space case.  This means
that in the special case of an abelian group one should recover the
known results from the flat space case.  In particular, the chiral
currents of the WZW theory are the nonabelian generalisation of the
free bosonic currents, and the affine algebra \eqref{eq:affalg}
satisfied by these currents is the natural nonabelian generalisation
of the algebra satisfied by the free bosonic fields \eqref{eq:oscalg}.

In this framework, it appears natural to consider the nonabelian
generalisation of the Neumann and Dirichlet conditions written down in
the previous section; we will therefore briefly discuss them below.
However, we should stress from the very beginning that these gluing
conditions for the chiral currents generically are \emph{not} the same
as the Neumann and Dirichlet boundary conditions for a string moving
in a group manifold.  The detailed derivation of the boundary
conditions from the algebraic gluing conditions will follow in the
next section.  We hope however that the following discussion will shed
some light on a few controversial statements in the literature.

\subsection*{Closed string}

Since the nonabelian analogues of $\d X(z)$ and $\bar\d X(\bar z)$ are
$-J_a(z)$ and $\bar J_a(\bar z)$, respectively, one can immediately
generalise the Neumann and Dirichlet conditions for free fields
written in terms of the one-forms in the last row in
Table~\ref{tab:freeND}.  One thus obtains
\begin{align*}
(J_a(z) dz + \bar J_a(\bar z) d\bar z)|B\rangle &= 0~,\tag{`N'}\\
(J_a(z) dz - \bar J_a(\bar z) d\bar z)|B\rangle &= 0~.\tag{`D'}
\end{align*}
For obvious reasons, as well as for convenience, we will refer to the
above conditions as `Neumann' and `Dirichlet' gluing conditions,
respectively.  In terms of the modes, these yield the following
\begin{align*}
(J_{an} - \bar J_{a~-n})|B\rangle &= 0~,\tag{`N'}\\
(J_{an} + \bar J_{a~-n})|B\rangle &= 0~.\tag{`D'}
\end{align*}
Notice that these conditions have an opposite relative sign with
respect to the ones written down in \cite{KO} (see also \cite{I}).
This discrepancy however is not related to the geometric
interpretation of these gluing conditions, as implied in \cite{AS},
but rather lies in the definition of the conserved currents with which
one starts \cite{Gaw}.  It must be said that the relative minus sign
in the definition of the chiral currents \eqref{eq:cc} is often
neglected in the literature without major consequences; nevertheless,
whenever the relation between the holomorphic and the antiholomorphic
sectors of the theory comes into play, this sign becomes crucial.

Alternatively, we can write these gluing conditions on the cylinder
where, by using the mode expansion
\begin{equation*}
J_a(w) = \sum_{n\in\Z}~J_{an}~e^{-nw}~,\qquad\qquad
\bar J_a(\bar w) = \sum_{n\in\Z}~\bar J_{an}~e^{-n\bar w}~,
\end{equation*}
and the fact that the boundary is at $w=-\bar w$, we obtain
\begin{align*}
(J_a(w) - \bar J_a(\bar w))|B\rangle &= 0~,\tag{`N'}\\
(J_a(w) + \bar J_a(\bar w))|B\rangle &= 0~.\tag{`D'}
\end{align*}

The same consistency conditions as in the case of the free theory
yield in this case nontrivial restrictions on the type of gluing
conditions we can impose in different directions.  Indeed, if we
impose the same type of gluing conditions in two different directions,
say $a$ and $b$, we must then have `Dirichlet' gluing conditions in
the directions $c$ of non-vanishing ${f_{ab}}^c$ since
\begin{equation*}
[J_{an} \mp \bar J_{a~-n},J_{bm} \mp \bar J_{b~-m}]|B\rangle =
{f_{ab}}^c(J_{c~n+m} + \bar J_{c~-(n+m)})|B\rangle~.
\end{equation*}
In particular, we cannot impose `Neumann' gluing conditions in all
directions unless the group is abelian.  Moreover, from
\begin{align*}
[J_{an} + \bar J_{a~-n},J_{bm} - \bar J_{b~-m}]|B\rangle =
&{f_{ab}}^c(J_{c~n+m} - \bar J_{c~-(n+m)})|B\rangle\\
&+ 2kn\delta_{n+m,0}~G_{ab}|B\rangle~, 
\end{align*}
we deduce that we can only have different gluing conditions in
orthogonal directions (that is, for which the corresponding element
$G^{ab}$ vanishes), and in that case we must have `Neumann' conditions
in the directions $c$ of non-vanishing ${f_{ab}}^c$.

\subsection*{Open string}

In this case the nonabelian generalisation of the Neumann and
Dirichlet gluing conditions reads
\begin{align*}
\left.(J_a(z) + \bar J_a(\bar z))\right|_{z=\bar z} &= 0~,\tag{`N'}\\
\left.(J_a(z) - \bar J_a(\bar z))\right|_{z=\bar z} &= 0~.\tag{`D'}
\end{align*}
On the modes we obtain
\begin{align*}
J_{an} + \bar J_{an} &= 0~,\tag{`N'}\\
J_{an} - \bar J_{an} &= 0~.\tag{`D'}
\end{align*}
Finally, on the cylinder, where the boundary is at $w=\bar w$, we
obtain the following gluing conditions
\begin{align*}
\left.(J_a(w) + \bar J_a(\bar w))\right|_{w=\bar w} &= 0~,\tag{`N'}\\
\left.(J_a(w) - \bar J_a(\bar w))\right|_{w=\bar w} &= 0~.\tag{`D'}
\end{align*}

We remark that in the open string picture there are no consistency
conditions constraining the allowed gluing conditions.  If one however
is interested in preserving the affine symmetry of the bulk theory
then one must keep in mind that, unlike the case of the free open
string, here it is only the `Dirichlet' gluing conditions that
preserve the affine algebra.  In other words, if we want to preserve
the current algebra, we cannot impose `Neumann' gluing conditions in
all directions.

This concludes our discussion of the generalisation of the standard
Neumann and Dirichlet gluing conditions to the group manifold case.
We can now consider the general case, and from now on we will work in
the open string picture.  Given a string background described by a WZW
model with target space $\mathbf{G}$ we consider the following gluing
conditions, where from now on we are implicitly evaluating both sides
of the condition at the boundary $z=\bar z$
\begin{equation}
J(z) = R{\bar J}(\bar z)~,\label{eq:GC} 
\end{equation}
where $R$ is a linear invertible map $R:\gg\to\gg$, with
$R(T_a)=T_b{R^b}_a$, for any $T_a$ in $\gg$.  The basic requirement
imposed on $R$ is that the resulting D-brane configurations be
conformally invariant.  This consistency condition imposes that the
holomorphic and antiholomorphic sectors agree at the boundary, that is
\begin{equation*}
\sT(z) = \bar\sT(\bar z)~.\label{eq:Gci}
\end{equation*}
Using the expression of the energy momentum tensor \eqref{eq:GT} we
obtain that the map $R$ must preserve the metric $\Omega$:
\begin{equation}
R^T \Omega R = \Omega~.\label{eq:Rmet}
\end{equation}
In the remaining of this paper we will implicitly assume that the
matrix of gluing conditions $R$ satisfies this condition.

\section{D-branes on group manifolds in the open string sector}

One of the most subtle issues of this approach is extracting the
geometric information from a given set of gluing conditions.  The
precise statement of the problem is the following: given a set of
gluing conditions \eqref{eq:GC} for the chiral currents and a fixed
but otherwise arbitrary point $g$ in the target group manifold
$\mathbf{G}$, find the possible D-branes which pass through $g$ and
are described by these gluing conditions.

It is clearly desirable to interpret the boundary conditions on group
manifolds in a similar manner with the flat space case; after all, the
special case when $\mathbf{G}$ is abelian is a flat space.  However,
in spite of the fact that the gluing conditions \eqref{eq:GC} have
been introduced as the nonabelian generalisation of
\eqref{eq:flatspaceBC}, there is an important distinction between
them.  In the case of a free theory the gluing conditions are formally
identical with the boundary conditions defined in the tangent space of
the target manifold; therefore the corresponding eigenvalues and
eigenvectors of the matrix $R$ identify the Neumann and Dirichlet
directions.  In the group manifold case the gluing conditions
\eqref{eq:GC} take values in the tangent space of $\mathbf{G}$
\emph{at the identity}, that is $T_e\mathbf{G}\equiv\gg$.  This is due
to the fact that the currents $J$ and $\bar J$ are themselves Lie
algebra valued objects.  Hence, in order to extract some geometric
information out of the algebraic gluing conditions we must first of
all `translate' them into boundary conditions in $T_g\mathbf{G}$, and
then determine what the Neumann and Dirichlet directions are in this
case.

Let us consider the left- and right-invariant Maurer--Cartan forms on
$\mathbf{G}$, which we denote by $\theta_L$, $\theta_R$.  We can then
use the maps $g:\Sigma\to\mathbf{G}$ to pull these Maurer--Cartan
forms back on $\Sigma$, thus obtaining
$g^*\theta_L = g^{-1}dg~$ and $g^*\theta_R = dg~g^{-1}$, which are
$\gg$-valued 1-forms that is, they belong to
$\Omega^1(\Sigma)\otimes\gg$.  We therefore see that the holomorphic
current $J(\sigma,\tau) = -\d g g^{-1}$ takes values in
$T^*\Sigma^{(1,0)}\otimes\gg$, whereas the antiholomorphic current
$\bar J(\sigma,\tau) = g^{-1}\bar\d g$ takes values in
$T^*\Sigma^{(0,1)}\otimes\gg$.

In order to obtain from the Lie algebra valued gluing condition
\eqref{eq:GC} a boundary condition taking values in $T_g\mathbf{G}$,
we need to `translate' the chiral currents from $T^*\Sigma\otimes\gg$
to $T^*\Sigma\otimes T_g\mathbf{G}$.  This is done with the help of
the two well-known maps, obtained from the left- and right-translation
in the group $G$
\begin{align*}
(\lambda_g)_* :\gg\to T_g\mathbf{G}~,\qquad\quad &
(\rho_g)_* :\gg\to T_g\mathbf{G}~,\\
(\lambda_g)_* X= g X~, \qquad\quad & (\rho_g)_* X = Xg~,
\end{align*}
which send elements of the Lie algebra into tangent vectors at the
point $g$.  If we now apply $(\rho_g)_*$ to both sides of the
gluing condition \eqref{eq:GC} we obtain the following boundary
condition in $T_g\mathbf{G}$:
\begin{equation*}
\d g =  \mathbf{R}(g) \bar\d g~,
\end{equation*}
where $\mathbf{R}(g)$ is the map $\mathbf{R}(g): T_g\mathbf{G} \to
T_g\mathbf{G}$, defined as
\begin{equation*}
\mathbf{R}(g) = - (\rho_g)_* \circ R \circ (\lambda_g)_*^{-1}~.
\end{equation*}
Thus, for any given tangent vector $V$ in $T_g\mathbf{G}$, we have
$\mathbf{R}(g)(V) = - R(g^{-1}V)g$.  First of all we must remark that,
by contrast with the flat space case, the matrix of boundary
conditions $\mathbf{R}(g)$ is \emph{not} the same as the matrix of
gluing conditions $R$, and the former is point-dependent.  
In general, for an arbitrary group $\mathbf{G}$, $\mathbf{R}$ only agrees
with $R$ at the identity (more generally, for any $g$ in the centre).
This implies, in particular, that the identification of the Neumann
and Dirichlet boundary conditions used in \cite{STDNW} can only find
D-branes which pass through the identity in the group manifold.

We can now safely identify the Neumann and Dirichlet directions.  In
the same way as in flat space, these are determined by the eigenvalues
and eigenvectors of the corresponding matrix of boundary conditions,
in our case $\mathbf{R}(g)$.  At a given point $g$, a Dirichlet
boundary condition corresponds to a $-1$ eigenvalue of
$\mathbf{R}(g)$, which means that the directions normal to the
worldvolume of the D-brane are spanned by the corresponding
eigenvectors of $\mathbf{R}(g)$.  All the other eigenvalues describe
Neumann boundary conditions and the corresponding eigenvectors span
the tangent space of the worldvolume of the D-brane.

Let us conclude this section by describing an alternative way of
deriving the boundary conditions in group manifolds, which proves to
be useful in concrete applications.  Given a particular group manifold
$\mathbf{G}$, we can parametrise it by introducing the coordinates
$X^{\mu}$, with $\mu=1,...,\dim\mathbf{G}$; we also introduce the
left- and right-invariant vielbeins defined by
\begin{equation*}
g^{-1}dg = {e^a}_{\mu}~dX^{\mu}T_a~,\qquad\qquad
dg g^{-1} = {\bar e}^a{}_{\mu}~dX^{\mu}T_a~.
\end{equation*}
These vielbeins are related by ${\bar e}^a{}_{\mu} = {e^b}_{\mu}
{A^a}_b$, where $A$ denotes the adjoint action of the group, $g T_a
g^{-1} = {A^b}_a T_b$.  Using this set-up, one can show that the
gluing conditions \eqref{eq:GC} give rise to the following boundary
conditions for the component fields $X^{\mu}$:
\begin{equation}
\d X^{\mu} = {\tilde R(g)}^{\mu}{}_{\nu} \bar\d X^{\nu}~.
\end{equation}
Here, the matrix of boundary conditions $\tilde R(g)$ is given by
\begin{equation}
\tilde R(g) = -{\bar e}^{-1} R e~.
\end{equation}
The Neumann and Dirichlet directions are identified in a similar way
as before.  Moreover, one can see once again that the matrix which
describes the boundary conditions at a given point in the target space
depends on that point, in this case, through the invariant vielbeins.

Our analysis so far has not made use of any particular property that
the map $R$ may have.  In the remaining of these notes we will
concentrate on two special classes of gluing conditions, for which a
detailed analysis of the resulting D-brane configurations is possible.

\section{Type-D gluing conditions}

The defining property of this type of gluing conditions is the fact
that the map $R$ is taken to be a Lie algebra automorphism, that is
\begin{equation}
[R(T_a),R(T_b)] = R([T_a,T_b])~.\label{eq:Rhom}
\end{equation}
This class can be thought of as a generalisation of D gluing
conditions, where the identity is replaced by a Lie algebra
automorphism 
\begin{equation}
J_a(z) - {R^b}_a {\bar J}_b(\bar z) = 0~.\label{eq:Dbc} 
\end{equation}
They are in some sense the most natural conditions, in that they
preserve the infinite-dimensional symmetry of the current algebra.
Indeed, from \eqref{eq:Rhom} and \eqref{eq:Rmet} it follows that $R$
also preserves the metric $G$ and hence the affine algebra
\eqref{eq:affalg}.

Let us now consider the geometry of the D-brane configurations they
describe.  If we start with the simplest case, where $R$ is taken to
be the identity matrix, then the corresponding matrix of boundary
conditions will be given by $\mathbf{R}(g) = - \Ad_{g^{-1}}$, and the
resulting D-branes can be identified, as shown in \cite{AS}, with the
conjugacy classes of the group $\mathbf{G}$.  Indeed, in this case,
provided that the metric $G$ restricts nondegenerately to the
conjugacy class $C$ of $g$, the tangent space at $g$ splits into the
tangent space to the conjugacy class and its perpendicular complement,
which can be identified with the tangent space to the centraliser
subgroup $Z$ of $g$:
\begin{equation*}
T_g\mathbf{G} = T_g C \oplus T_g Z\quad\text{with}\quad T_gC \perp T_g 
Z~.
\end{equation*}
Moreover $\Ad_{g^{-1}}$ restricts to the identity on $T_gZ$, which
means that the Dirichlet directions span $T_g Z$.  Furthermore, the
Neumann directions span $T_g C$, and hence the worldvolume of the
D-brane can be identified with $C$.

As the next step, let us now consider $R$ to be an inner
automorphism.  It therefore can be identified with $\Ad_r$, for some
group element $r$, and consequently the gluing conditions
\eqref{eq:Dbc} become
\begin{equation}\label{eq:rGC}
-\d g g^{-1} = r g^{-1} \bar\d g r^{-1}~.
\end{equation}
If we introduce a new field $\tilde g = gr^{-1}$, we can write the
corresponding boundary conditions in the following form
\begin{equation*}
\d\tilde g = -\Ad_{\tilde g^{-1}} \bar\d\tilde g~.
\end{equation*}
By using the previous argument, applied this time to $\tilde g$, we
are lead to conclude that the corresponding D-brane configuration,
which is described by the field $g$, has a worldvolume which lies
along the right--translate $Cr$ of the conjugacy class of $g$ by the
element $r$.  This result differs from the one obtained in \cite{AS},
where it was argued that inner automorphisms, being symmetries of the
model, cannot result in D-brane configurations different from the the
ones already described by $R=\1$.  Although inner automorphisms are
symmetries of the string background, they are not necessarily
symmetries of the theory containing a D-brane.  This fact does not
constitute a novelty, as D-branes break some of the bulk symmetries
even in flat space (e.g., translational symmetry).

One remark is in order.  One can alternatively apply $\Ad_{r^{-1}}$ in
both sides of the gluing condition \eqref{eq:rGC} and obtain, using a
similar argument, that the corresponding D-brane lies along the
left-translate $rC$ of the conjugacy class of $g$.  Since however
$Cr=rC$, this does not lead to any ambiguity.

It is interesting to point out that conjugacy classes in groups
admitting bi-invariant metrics have always even dimension.  Indeed,
conjugacy classes are the image under the exponential map of adjoint
orbits which are diffeomorphic to co-adjoint orbits.  Co-adjoint
orbits, on the other hand, are symplectic manifolds relative to the
natural Kirillov--Kostant--Souriau symplectic structure and hence are
even-dimensional.  Therefore the D-branes whose worldvolumes can be
identified with shifted conjugacy classes have even-dimensional
worldvolumes.

We now turn to the case of a general automorphism.  We would like to
identify the submanifold $N$ of $\mathbf{G}$ that corresponds to the
worldvolume of the D-brane described by \eqref{eq:Dbc}.  Provided the
metric on $\mathbf{G}$ restricts nondegenerately to $N$, we can split
the tangent space $T_g\mathbf{G}$ into the tangent space to $N$ and
its orthogonal complement:
\begin{equation*}
T_g\mathbf{G} = T_g N \oplus T_g N^{\perp}~.
\end{equation*}
Let us consider a given but otherwise arbitrary vector $V$ in $T_g
N^{\perp}$.  As we saw in the previous section, any vector $V$ normal,
at a point $g$, to the worldvolume of the D-brane is an eigenvector of
the matrix $\mathbf{R}(g)$ corresponding to an eigenvalue equal to
$-1$.  It therefore satisfies
\begin{equation*}
R(g^{-1}V) = Vg^{-1}~,
\end{equation*}
which in turn implies that
\begin{equation*}
\langle R(g^{-1}V) - Vg^{-1}, R(X)\rangle = 0~,
\end{equation*}
for any element $X$ in the Lie algebra.  From this relation, by using
the fact that $R$ preserves the metric, as well as the fact that the
metric is bi-invariant, we can deduce that
\begin{equation*}
\langle V, R(X)g - gX \rangle = 0~.
\end{equation*}
This relation tells us that the vector $W_X\equiv R(X)g - gX$, which
belongs to $T_g\mathbf{G}$, is normal to $V$, for any $X$.  Hence
$W_X$ is tangent to $N$ that is, to the worldvolume of the D-brane.

If $R=\1$, then $W_X$ would simply be a vector tangent to the
conjugacy class $C$ of $g$.  In general, $W_X$ turns out to be tangent 
to a `twisted' conjugacy class, which is defined as follows (for more
details, see the Appendix)
\begin{equation}
C_R = \left\{ r(h)gh^{-1} \mid h\in\mathbf{G}\right\}~,
\end{equation}
where the map $r:\mathbf{G} \to \mathbf{G}$ is defined by
\begin{equation*}
r\left(e^{tX}\right) = e^{tR(X)}~,
\end{equation*}
for $t$ small enough and $X$ any element in the Lie algebra.  

In order to see that $W_X$ is tangent to $C_R$ we consider a curve in
$C_R$ 
\begin{equation*}
\gamma_R(t) = r(h(t))gh^{-1}(t)~,
\end{equation*}
which is defined in terms of a particular curve $h(t)$ in
$\mathbf{G}$, characterised by $h(0)=\1$ and $\dot h(0) = X$.  This
curve passes through $g$ and its tangent vector at $g$ is nothing but
$W_X$ 
\begin{equation*}
\left.\frac{d\gamma_X(t)}{dt}\right|_{t=0} = R(X)g - gX~.
\end{equation*}
Hence for any Lie algebra element $X$ there exists a tangent vector
$W_X$ to the D-brane, which is also tangent to the twisted conjugacy
class $C_R$ of $g$.  If we could now show that any tangent vector to
the D-brane is of the form $W_X$, for some $X$ in $\gg$, then we would
be able to conclude that the corresponding D-brane can be identified
with $C_R$.

We know that $T_g N^{\perp}$ is spanned by the Dirichlet eigenvectors
of $\mathbf{R}(g)$.  One can rephrase this by saying that any vector
$V$ in $T_g N^{\perp}$ belongs to the kernel of the following operator
\begin{equation*}
\1 - (\rho_g)_* \circ R \circ (\lambda_g)_*^{-1}~.
\end{equation*}
Therefore $T_g N$, which is the orthogonal complement of $T_g
N^{\perp}$, is nothing but the image of the adjoint of the above
operator which is given by
\begin{equation*}
\left(\1 - (\rho_g)_* \circ R \circ
(\lambda_g)_*^{-1}\right)^{\dagger} = \1 - (\lambda_g)_* \circ R \circ
(\rho_g)_*^{-1}~.
\end{equation*}
In order to obtain the above relation we used the fact that
$(\rho_g)_* \circ R \circ (\lambda_g)_*^{-1}$ is an isometry and
therefore its adjoint is given by its inverse.  This further implies
that any vector $W$ in $T_g N$ can be written as
\begin{equation*}
W = U - gR^{-1}(Ug^{-1})~,
\end{equation*}
for some vector $U$ in $T_g\mathbf{G}$.  Since, moreover, every such
vector $U$ in $T_g\mathbf{G}$ satisfies $Ug^{-1} = R(X)$, for some Lie
algebra element $X$, it follows that $W=R(X)g-gX$, for some $X$.  We
can therefore conclude that the worldvolume of the D-brane can be
identified with the twisted conjugacy class $C_R$.

Until now we have assumed that the worldvolume of the D-brane forms a
submanifold $N$ of $\mathbf{G}$.  In order to check that this is
indeed the case, one must show that the tangent vectors to $N$ satisfy
the Frobenius integrability condition, that is, $[W_1(g),W_2(g)]$
belongs to $T_g N$ whenever $W_1(g)$ and $W_2(g)$ do.  If we compute
this bracket explicitly in our case we obtain
\begin{equation*}
[R(X)g - gX, R(Y)g - gY] = R([XY])g - g[XY]~,
\end{equation*}
where we have used the fact that $R$ is an automorphism.  Hence D-type
gluing conditions always give rise to D-branes which are submanifolds
of $\mathbf{G}$.  This implies, in particular, that such gluing
conditions cannot describe configurations consisting of intersecting
D-branes. 

Notice that in this case we do not necessarily obtain odd-dimensional
D-branes, as the relation between conjugacy classes and co-adjoint
orbits endowed with a symplectic structure cannot be generalised to
the `twisted' case.  In fact, in this case, one can find explicit
solutions describing even-dimensional D-branes \cite{FSDNW}.

\section{Type-N gluing conditions}

While the particular type of gluing conditions discussed in the
previous section is in some sense the most natural, as it preserves
the current algebra of the bulk theory, it is certainly not the most
general.  In this section we consider a different type of gluing
conditions which is defined in terms of the Lie algebra automorphism
\eqref{eq:Rhom}.  This case, however, can be thought of as a
generalisation of the `Neumann' gluing conditions, being defined as follows
\begin{equation}
J_a(z) + {R^b}_a {\bar J}_b(\bar z) = 0~.\label{eq:Nbc} 
\end{equation}
The coefficients ${R^b}_a$ are, as before, the matrix elements of
the map $R$, with $R(T_a)=T_b{R^b}_a$, for any $T_a$ in $\gg$.  Notice
that here, although the holomorphic and antiholomorphic sectors are
related by a metric preserving automorphism, the infinite-dimensional
symmetry of the current algebra is \emph{not} preserved, contrary to a
statement made in \cite{STDNW}.  

In this case we do not yet understand the general picture of the
possible D-brane configurations that one can obtain.  For this reason, 
we will restrict ourselves to considering a concrete example, which is
the case of the group $\mathbf{G} = \SU(2)$.  It is convenient to
choose a parametrisation for $\SU(2)$
\begin{equation}\label{eq:parg}
g = e^{\theta_2 T_2}e^{\theta_1 T_1}e^{\theta_3 T_3}~,
\end{equation}
where $\theta_1,\theta_2,\theta_3$ play the r\^ole of the spacetime
fields, $\{T_1,T_2,T_3\}$ forms a basis for $\mathfrak{su}(2)$, and
the brackets are $[T_aT_b]=\epsilon_{abc}T_c$.
The left- and right-invariant vielbeins read
\begin{equation*}
e = \begin{pmatrix}
    \cos\theta_3  & \cos\theta_1\sin\theta_3 & 0\\
    -\sin\theta_3 & \cos\theta_1\cos\theta_3 & 0\\
                0 &            -\sin\theta_1 & 1
    \end{pmatrix},\qquad
\bar e = \begin{pmatrix}
         \cos\theta_2 & 0 & \cos\theta_1\sin\theta_2\\
                    0 & 1 & -\sin\theta_1\\
        -\sin\theta_2 & 0 & \cos\theta_1\cos\theta_2
         \end{pmatrix}.
\end{equation*}
Notice, first of all, that this parametrisation is singular whenever
$\cos\theta_1=0$.  One can in fact show that these `singular' points
give rise to two $S^1$'s inside $S^3$, described by
\begin{equation*}
g_2(\pm\pi/2,\phi_2,\phi_3) = e^{\pm\pi/2 Y_1}
                              e^{(\phi_3\mp\phi_2)Y_3}~.
\end{equation*}

Let us start with the simplest case, $R=\1$, for which the gluing
conditions read
\begin{equation}\label{eq:NR=1}
J(z) = - \bar J(\bar z)~.
\end{equation}
This case can be thought of as the nonabelian generalisation of the
flat space case where we have Neumann boundary conditions in all
directions.  As is well-known, in the flat space case one obtains
D-branes whose worldvolume fills the whole target; here the possible
D-branes are described by the boundary conditions that \eqref{eq:NR=1}
give rise to, namely
\begin{equation}
\d\theta^{\mu} = {\tilde R}(g)^{\mu}{}_{\nu}\bar\d\theta^{\nu}~,
\end{equation}
where $\tilde R(g) = \bar e^{-1}e$.  If we evaluate the matrix of
boundary conditions at the identity we obtain $\tilde R(\1)=\1$, which
indicates that we have three $+1$ eigenvalues and hence three Neumann
directions spanning $\mathfrak{su}(2)$.  In other words, the identity
in $\SU(2)$ belongs to an euclidean D-brane having a three-dimensional
worldvolume.  As soon as we move away from the identity, $\tilde R$
will will no longer be $\1$.  In general, $\tilde R$ will have one
$+1$ and two complex conjugate eigenvalues.  Therefore, at generic
points in the group manifold, there will be no $-1$ eigenvalues and
thus no Dirichlet directions.  Nevertheless, there exist submanifolds
of $\SU(2)$ where $\tilde R$ has at least one -1 eigenvalue; in fact,
since $\det\tilde R=1$, it will necessarily have two such eigenvalues.
These submanifolds can be described as the zero locus of a function,
$F(g)\equiv\Tr\tilde R(g) + 1$; in our parametrisation this is given
by
\begin{equation*}
F = 1 + \cos\theta_1\cos\theta_2 + \cos\theta_2\cos\theta_3 +
      \cos\theta_3\cos\theta_1 +
      \sin\theta_1\sin\theta_2\sin\theta_3~.
\end{equation*}
Let us denote by $\eF$ the zero locus of $F$.  Since every point in
$\eF$ is characterised by one $+1$ and two $-1$ eigenvalues of $\tilde
R$, it follows that through every point on $\eF$ passes an euclidean
D-particle, having as the tangent vector to its worldline the
eigenvector of $\tilde R$ corresponding to the $+1$ eigenvalue.  Two
natural questions arise.  First, we would like to have a better
geometric understanding of the surface $\eF$.  Second, we have to
determine whether or not these particular solutions are consistent; in
other words we must make sure that their worldlines lie on the surface
$\eF$.

In order to answer the first question we consider the vector fields
that generate the adjoint action of a group $\mathbf{G}$.  As it is
shown in the Appendix, these vector fields can be written in terms
of the right- and left-invariant vielbeins as follows
\begin{equation}\label{H_a}
H_a(g) = \left( ({\bar e}^{-1})^{\mu}{}_a - (e^{-1})^{\mu}{}_a
\right) \d_{\mu}~,\qquad\qquad a=1,2,3.
\end{equation}
It is now a straightforward calculation to check that that, in our
case, $F$ is annihilated by the corresponding vector fields $H_a$
\begin{equation*}
H_a(g)\cdot F(g) = 0~,\qquad a=1,2,3.  
\end{equation*}
This shows that $F$ is a class function on $\SU(2)$, that is
$F(hgh^{-1})=F(g)$, for any group elements $g$ and $h$.  From this we
can conclude that the zero locus of $F$ consists of adjoint
orbits---that is, conjugacy classes.

Let us now turn to the second question.  Since $\tilde R$ only takes
$-1$ eigenvalues on the surface $\eF$, it means that if we have a
solution describing a D-particle its worldline should lie the above
surface.  In order to check that this is indeed the case we must show
that the vector tangent to the worldline is also tangent to $\eF$.  If
we compute explicitly the Neumann eigenvector $V_1$ of $\tilde R$, we
can easily check that
\begin{equation*}
\left. V_1(g)\cdot F(g)\right|_{F(g)=0} = 0~.
\end{equation*}
In other words, $V_1$ is tangent to the zero locus of $F$, that is
$\eF$.

Let us summarise our findings so far.  We have seen that the gluing
conditions \eqref{eq:NR=1} give rise two types of euclidean D-branes.
Around the identity we have a D2-brane which extends up to a surface
defined as the zero locus of a class function.  Moreover, this surface
itself is foliated by the worldlines of D-particles. 


Let us now consider the gluing conditions \eqref{eq:Nbc} with an
arbitrary $R$.  Since $\mathfrak{su}(2)$ is a simple Lie algebra, $R$
is an inner automorphism; hence it can be identified with $\Ad_r$, for
some group element $r$.  By following the same type of argument as in
the case of the D-type gluing conditions, one can show that also in
this case the effect of an inner automorphism at the level of the
gluing conditions is a translation in the group manifold.  Indeed, now
the matrix of boundary conditions $\tilde R=\bar e^{-1}R e$ gives rise
to three Neumann eigenvectors everywhere except a submanifold which we
denote by $\eF_r$.  This submanifold is identified as the zero locus
of a function $F_r(g) = \Tr\tilde R(g) +1$.  Clearly, this function is
related to the previous $F$, as it satisfies
\begin{equation*}
F_r(g)=F(gr^{-1})~.
\end{equation*}
Therefore $\eF_r$ is nothing but the translation of $\eF$ by the group
element $r$, that is $\eF_r=\eF r$.  Hence through every point in $\eF
r$ passes a D-particle whose worldline lies on $\eF r$.  Moreover, not
only the surfaces $\eF$ and $\eF_r$ are related by translation, but so
are the Neumann eigenvectors tangent to the worldlines of the
corresponding D-particles on these surfaces.

A particularly interesting case is the one where $R$ itself has two
$-1$ eigenvalues.  In this case, the corresponding surface $\eF r$
passes through the identity element in $\SU(2)$; therefore there 
exists a particular D-particle whose worldline passes through the
identity.  Its tangent vector, at the identity, is simply given by 
$V_1=\d_{\theta_3}$, as expected.  The worldline of this D-particle is 
nothing but the subgroup of $\SU(2)$ generated by $T_3$.  Moreover,
the translation of this particular solution gives rise to D-particle
configurations whose worldlines are cosets in $\SU(2)$.





\section{Conclusions and outlook}

In this paper we have analysed the geometric interpretation of
D-branes in group manifolds obtained via the boundary state formalism.
Although the algebraic analysis of the possible gluing conditions is
quantum, the subsequent analysis of the geometry of the resulting
D-branes is essentially classical.  (This situation is somewhat
similar to the analysis in \cite{OOY,BBMOOY} of geometry of D-branes
in Calabi-Yau manifolds which was carried out in the large volume
limit.)  The picture that emerges is the following.  The algebraic
gluing conditions that one usually imposes on the chiral currents of
the WZW model are not to be understood as boundary conditions in the
target space; however the latter can be derived from the former by
an appropriate `translation' in the group manifold.  The boundary
conditions that one obtains in this way are point dependent, by
contrast to the more familiar case of the flat space.

Determining the D-brane configurations that the most general gluing
conditions give rise to is a relatively difficult task.  Here we have
analysed in detail the simplest case of gluing conditions, which
preserve the affine symmetry of the bulk theory.  The so-called type-D
gluing conditions give rise to D-branes whose worldvolumes lie on
twisted conjugacy classes.  These can be understood as a
generalisation of the familiar notion of a conjugacy class, which
depends on the automorphism defining the gluing conditions.  In the
particular case of inner automorphisms we obtain D-branes with
even-dimensional worldvolumes.  We have also considered a second class
of gluing conditions, which we called type-N.  They give rise to more
complicated configurations which, at least in a particular case,
include subgroups, cosets and open submanifolds of dimension equal to
the dimension of the target space.

The study carried out in this paper was restricted to the case where
the gluing conditions imposed on the affine currents are given in
terms of a constant map $R$.  This implicit restriction has no
conceptual basis, it is merely a practical one, dictated by our
inability of working at the level of the quantum theory with field
dependent quantities.  Field dependent gluing conditions do however
appear \cite{STDNW} in the study of the boundary conditions coming
from the classical sigma model action of the the WZW model.  It would
therefore be interesting to consider a generalisation of the formalism
presented here in this particular direction.

Here we have taken the point of view that the WZW model is a typical
example of an exact string background whose CFT is known explicitly
(at least the generators of the conformal (super)algebra); therefore
the natural approach to the study of the possible D-brane
configurations is to impose gluing conditions on the fields in terms
of which the CFT is defined, so that the requirement of conformal
invariance can be easily implemented.  This is however only one side
of the story.  The WZW model provides also a typical example of a
string background admitting a sigma model description. This allows one
to undertake a complementary study of the possible D-brane
configurations \cite{KlS,Gaw}.  Furthermore, one must address the
question concerning the relation between the two approaches
\cite{STDNW}.  In this context it is perhaps worth remarking that the
results obtained here, in particular the two types of D-brane
configurations in the case of $\SU(2)$, bear a certain similarity with
the ones obtained in \cite{KlS} from the analysis of the open string
WZW model.  Since the two classes of solutions obtained there are
related by Poisson--Lie T-duality, it would be interesting to
investigate such a possibility in the framework of the boundary state
approach.

\section*{Acknowledgements}

It is a pleasure to thank JM~Figueroa-O'Farrill and AA~Tseytlin for
many useful discussions and for a critical reading of the manuscript.
Some of the motivation for undertaking this work came from discussions
with AY~Alekseev and V~Schomerus to whom I extend in this way my
thanks.  This work was supported by a PPARC Postdoctoral Fellowship.

\section*{Appendix: (Twisted) conjugacy classes}

Let $\mathbf{G}$ be a connected Lie group and $g$ a fixed but
otherwise arbitrary element of $\mathbf{G}$.  Let
$C(g)$ denote the conjugacy class of the element $g$, defined as the
subset of $\mathbf{G}$ with the following elements:
\begin{equation*}
  C(g) := \left\{ h g h^{-1} \mid h \in \mathbf{G} \right\}~.
\end{equation*}
The conjugacy class of an element $g$ is therefore the orbit of that
element under the adjoint action of the group: $\Ad_h : \mathbf{G} \to
\mathbf{G}$, defined by $\Ad_h (g) = h g h^{-1}$.  Each conjugacy
class is a connected submanifold of $\mathbf{G}$.  Since every element
$g$ belongs to one and only one conjugacy class, $\mathbf{G}$ is
foliated by its conjugacy classes.
 
We have seen in Section 5 that it is useful to consider a certain
generalisation of the notion of a conjugacy class which is defined as
follows:
\begin{equation}
C_R(g) := \left\{ r(h)gh^{-1} \mid h\in\mathbf{G}\right\}~,
\end{equation}
where the map $r:\mathbf{G} \to \mathbf{G}$ is defined by
\begin{equation*}
r\left(e^{tX}\right) = e^{tR(X)}~,
\end{equation*}
for $t$ small enough and $X$ any element in the Lie algebra.  If we
assume that $\mathbf{G}$ is connected then $r$ extends to a Lie group
automorphism.  Moreover $r$ preserves the bi-invariant metric on the
Lie group.  The twisted conjugacy class of a given element is
therefore the orbit of that element under the action of $\Ad^R_h :
\mathbf{G} \to \mathbf{G}$, defined by $\Ad^R_h (g) = r(h) g h^{-1}$.

A curve in $C_R(g)$ is generically given by $\gamma_R(t) =
r(h(t))gh^{-1}(t)$, with $h(t)$ an arbitrary curve in $\mathbf{G}$.
Let us now consider a particular type of curve in $C_R(g)$, such that
$h(0)=\1$, and $\dot h(0) = Y$, with $Y=Y^a T_a$ an element of the
corresponding Lie algebra $\gg$.  The tangent vector to this curve at
the point $g$ will be given by
\begin{equation*}
\left.\frac{d}{dt}\left(r(h(t))gh^{-1}(t)\right)\right|_{t=0} = R(Y)g
                                                                - gY~, 
\end{equation*}
which is an element of $T_g\mathbf{G}$.  This defines a vector field
$H^R$ given by
\begin{equation*}
H^R(g) = R(Y)_R(g) - Y_L(g)~,
\end{equation*}
where we denote by $Y_{L,R}$ the left- and right-invariant vector
fields corresponding to $Y$.  One can further show that the components
of these invariant vector fields, defined by 
\begin{equation*}
Y_L(g) = Y_L^{\mu}(g)\d_{\mu}~,\qquad\quad
R(Y)_R(g) = R(Y)_R^{\mu}(g)\d_{\mu}~,
\end{equation*}
are given by
\begin{equation*}
Y_L^{\mu}(g) = Y^a(e^{-1})^{\mu}{}_a~,\qquad\quad
R(Y)_R^{\mu}(g) = Y^a {R^b}_a({\bar e}^{-1})^{\mu}{}_b~.
\end{equation*}
This implies that the vector field $H^R(g)$ which is tangent, at every
point $g$ to the curve $\gamma_R(t)$ in $C_R(g)$ is given by
\begin{equation*}
H^R(g) = Y^a \left( ({\bar e}^{-1})^{\mu}{}_b {R^b}_a -
         (e^{-1})^{\mu}{}_a \right) \d_{\mu}~.
\end{equation*}
This allows us to define $\dim\gg$ vector fields $H^R_a$, corresponding
to each of the generators $T_a$ of $\gg$:
\begin{equation*}
H^R_a(g) = \left( ({\bar e}^{-1})^{\mu}{}_b {R^b}_a -
           (e^{-1})^{\mu}{}_a \right) \d_{\mu}~.
\end{equation*}
These vector fields thus generate the `twisted' adjoint action of the
group $\mathbf{G}$.

The particular case of the standard conjugacy class can be easily
obtained by setting $R=\1$.  The vector fields that generate the
adjoint action of $\mathbf{G}$ read
\begin{equation*}
H_a(g) = \left( ({\bar e}^{-1})^{\mu}{}_a - (e^{-1})^{\mu}{}_a \right)
         \d_{\mu}~. 
\end{equation*}

%
%

\providecommand{\href}[2]{#2}\begingroup\raggedright\endgroup

\end{document}